Title: Wolumes - An algorithm to compute the volume of atoms and residues in proteins
Authors: Oliviero Carugo (Department of Chemistry, University of Pavia, Pavia, Italy and Department of Structural and Computational Biology, Max F. Perutz Laboratories, Vienna University, Vienna, Austria)
Category: q-bio.BM

Wolumes is a fast and stand-alone computer program written in standard C that allows the measure of atom volumes in proteins. Its algorithm is a simple discretization of the space by means of a grid of points at 0.75 Angstroms from each other and it uses a set of van der Waals radii optimized for protein atoms. By comparing the computed values with distributions derived from a non-redundant subset of the Protein Data Bank, the new methods allows to identify atoms and residues abnormally large/small. The source code is freely available, together with some examples.

## 1 Introduction

Protein structure analysis is of crucial importance in molecular biology. Since the development of the first tools, like the Ramachandran map [1, 2], many other conceptual instruments and computer programs have been developed.

All of them have a dual potential: they can be used to inspect and extract biochemical information from structural data or to validate structural results [3]. For example, the analysis of cavities in the protein interior or at the protein surface may help in identifying ligand binding sites and may also indicate regions of anomalous and perhaps incorrect packing of the residues [4, 5].

Some of the protein structure analysis tools are based on chemical principles. For example the analysis of the network of hydrogen bonds is extremely important for understanding protein stability and flexibility and the mechanisms of protein function [6]. Other tools are on the contrary based only on geometrical analyses, for example the measures of solvent accessible surface area [7, 8] and atom burial degree [9, 10], where it is just necessary to fix the atom and/or the solvent probe dimension and no chemical and physical properties are included in the computations.

Contrary to solvent accessibilities, which are routinely computed in a myriad of applications that range from docking simulations to folding studies, atom and residue volumes received minor attention. However, atom and residue volumes are of interest to evaluate the packing in the core of globular proteins [11], to assess the quality of structural models [12], and for use in the implicit solvent models used, for example, in the generalized Born model of electrostatic solvation [13]

Atom and residue volumes can be computed numerically either by discretizing the space occupied by a protein into an equivalent ensemble of discontinuous subspaces or by a tessellation procedure that assigns a fraction of the total space to each atom or residue. For example, the volumes of the amino acids can be computed by means of a Laguerre tasselation with the server VLDP (www.dsimb.inserm.fr/dsimb_tools/vldp/) [14] while a discretization approach is used by the program ATVOL that computes atom volumes by separating neighbour atoms where their van der Waals spheres clashes and by performing numerical integrations to measure the volume of each atom (http://www.ctu.edu.vn/~dvxe/Bioinformatic/Software/kinemage/kinemage.biochem.duke.edu/software/software3.html#atvol).

The Voronoi polyhedra, which are the minimally sized polyhedra around each atom and allow the allocation of the space within the protein, including the interstices, to its constituent atoms, were used by Tsai and colleagues to estimate atom and residue volumes [15, 16]. Six, dissimilar sets of atom volumes were computed by Schaefer and colleagues, for three ensembles of parameters of the CHARMM software package [17] and with two alternative methods, one based on the Voronoi polyhedra and the other based on the superposition of individual atom volumes [18]. Voronoi decomposition and Delaunay triangulation were also implemented into a fast computer program able to detect cavities in large proteins and to measure their volume [19].

Despite numerical methods, like for example grid based or probe sphere detection techniques, tend to be less accurate than analytical methods, in the sense that they can reach high levels of accuracy only at an exorbitant computational cost, it is necessary to consider that the accuracy of the protein three-dimensional models resulting from crystallographic of NMR spectroscopy experiments is relatively modest. Therefore, also by considering the considerable flexibility of these molecules, which is not properly described by a static snapshot, it is questionable if the pursuit of a higher accuracy is not a sterile and useless exercise. A recently designed grid based methods for computing molecular volumes was applied to detect cavities and surface clefts in proteins and was not used to measure atom volumes [20].

These two approaches are different not only in their computational strategy but also in their conceptual bases. While with a tessellation approach the space is integrally occupied by atoms and no void interstices are considered to be possible, empty cavities are possible by measuring atom volumes with a discretization method.

Here a new program for measuring atom and residue volumes in proteins is described. It is based on a simple and fast discretization of the space and, although its results compare well with those of other computer programs, it is tens of times faster. Being a stand-alone program written in standard C, it can be used with any operating system independently of any other software licences.

## 2 The program

### 2.1. The algorithm

The protein structure is inserted into a three-dimensional grid of points (see Figure 1 for a bidimensional sketch). Each point is surrounded by an octahedron of six points, each at a distance D. The borders of the grid are determined by the dimension of the protein. The minimal value of x is fixed at xm-3 Å, where xm is the minimal value of x for the atoms of the protein; the maximal value of x is equal to xM+3 Å, where xM is the maximal value of x for the atoms of the protein; the borders of the y and z coordinate are selected analogously.

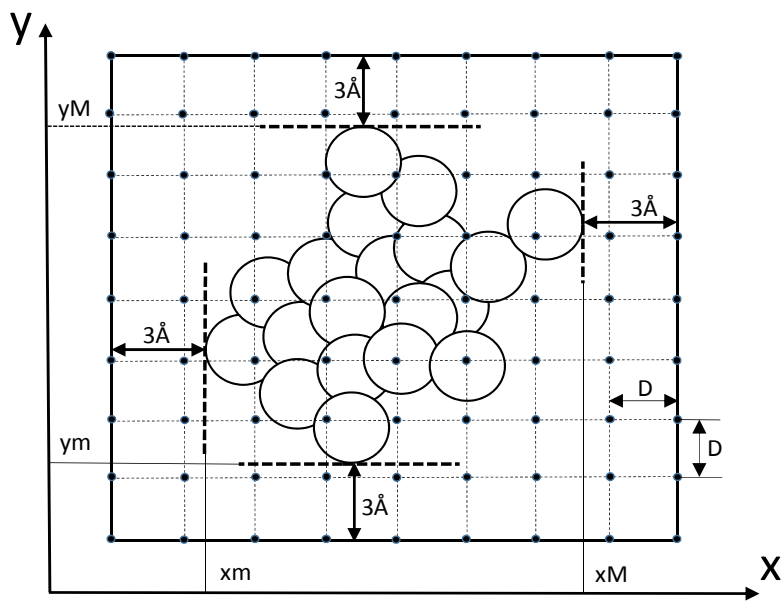

*Figure 1. Bidimensional example of a grid that surrounds a protein molecule. The atoms are shown as white circles, the grid points as small black circles. The borders of the grid depends on the dimension of the protein. The distance between the grid points is indicated with D.*

Each atom is characterized by its van der Waals radius (taken from [21]) and a point of the grid can fall either inside the van der Waals sphere of the atom, and in this case it belongs to this atom, or outside (Figure 2). If a point of the grid falls into the van der Waals sphere of two (or more) atoms, it is considered to belong to both (or to all) of them in equal parts. The volume of an atom is equal to the number (N) of points of the grid that belong to it. Since each point of the grid occupies a volume of $D^3$ Angstrom, the volume occupied by an atom is equal to N x $D^3$ $Å^3$.

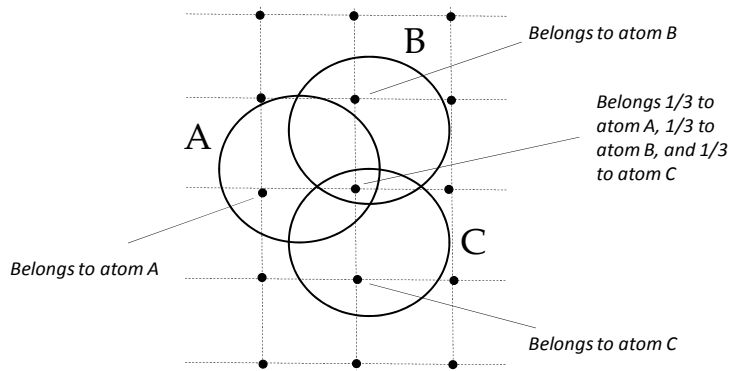

*Figure 2. Criteria used to measure the volume of an atom. The volume depends on the number of grid points that belong to the atoms. In the example, all the atoms (A, B, and C; shown as white circles) have the same volume, since each of them contains 4/3 of grid points (shown as small black circles).*

**2.2 Grid optimization**

Various values of D were tested. Small values obviously ensure more accurate measurements than large values. However, small values require longer computation time than large values. Consequently, it is necessary to mediate between the two requirements: accuracy and computational speed. Clearly, the computational speed depends severely on the grid dimension. The data shown in Table 1 were determined by computing the atom volumes of the PDB file 135l, which contains 994 atoms. They can be fitted with the equation *time* = 2.82 + $D^{-2.96}$ (correlation coefficient = 1.000) or with the equation *time* = $D^{-3.42}$ (correlation coefficient = 1.000).

*Table 1. CPU time necessary to process the PDB file 135l with the program WOLUMES at varying values of the variable D.*

| D (Å) | CPU time (seconds) |
|---|---|
| 0.10 | 2643.5 |
| 0.25 | 168.2 |
| 0.50 | 21.3 |
| 0.75 | 6.4 |
| 1.00 | 2.7 |
| 1.25 | 1.5 |
| 1.50 | 0.9 |
| 2.00 | 0.4 |
| 3.00 | 0.1 |

The dependence of the accuracy on D was analyzed by computing the volumes of the atoms of the PDB entry 135l with the following nine D values: 0.1, 0.25, 0.5, 0.75, 1, 1.25, 1.5, 2, and 3 Å. By assuming that the most accurate volume values can be computed with D = 0.1 Å, the variations from these values were computed when D > 0.1 Å. **Table 2** indicates that an accuracy larger them 95% is reached with D =

0.75 Å. On the contrary, an average mistake of 6.7% occurs if D = 1 Å and larger mistakes are done for larger D values. Therefore, it seems reasonable to use a D value of 0.75 Å in order to compute atom volumes. It might be interesting also to note that the data shown in Table 2 can be fitted with the equation Mistake = 0.28.D + 6.6.D2 (correlation coefficient = 0.999), which can be used to estimate the relative precision associated with the selection of alternative values of D.

Table 2. *Average mistakes of the atom volumes of the PDB file 135l computed with the program WOLUMES at varying values of the variable D. The mistakes are assumed to be equal to zero at D = 0.1 Å.*

| D (Å) | Average mistake % (standard error) |
|---|---|
| 0.10 | - |
| 0.25 | 0.4(0.1) |
| 0.50 | 1.6(0.1) |
| 0.75 | 4.2(0.1) |
| 1.00 | 6.7(0.2) |
| 1.25 | 10.8(0.3) |
| 1.50 | 13.4(0.3) |
| 2.00 | 29.2(0.7) |
| 3.00 | 59.9(1.5) |

**2.3 Speed**

The speed of this algorithm was estimated by computing the atom volumes on six PDB entries and by using the D value of 0.75 Å. From the data shown in Table 3 it appears that the dependence between computation time and number of atoms is not linear. The analytical relationship that can be used to predict the processing time is Time = 0.014 nato + 4.22 x $10^{-6}$ nato$^2$ (correlation coefficient = 1.000).

Table 3. *CPU time necessary to process the six PDB files of different dimension with the program WOLUMES.*

| PDB | CPU time (seconds) | Number of atoms, nato |
|---|---|---|
| 135l | 6 | 994 |
| 104m | 8 | 1217 |
| 3cba | 42 | 1844 |
| 2sod | 141 | 4368 |
| 4aah | 562 | 9894 |
| 9xim | 787 | 12188 |

**2.4 Comparison with other programs**

A set of 501 protein structures was extracted from the Protein Data Bank [22, 23] in order to compare the program WOLUMES with other programs that allow one to perform similar computations. These 501 structures were selected according to the following criteria: monomeric molecules with resolution not worse than 2 Å and pairwise sequence identity smaller than 30%, excluding membrane proteins, proteins where more than 5% of the atoms are heteroatomic (excluding water molecules), and proteins with missing atoms or residues.

The volumes computed with the new program have been compared with those computed by means of a Laguerre tasselation with the server VLDP, which provides the volumes of the residues (and not of the individual atoms) (www.dsimb.inserm.fr/dsimb_tools/vldp/) [14]. The residue volumes computed with WOLUME tend to be systematically smaller that those computed with VLDP, though there is an excellent linear correlation (correlation coefficient = 0.965). The absolute value of the differences between the residue volumes computed with VLDP and WOLUMES is equal, on average, to 24.7 (±0.9) Å$^3$. Given that there are about 7.5 atoms per residue in the proteins, this means that the mean difference between the atom volumes computed with these two programs is about 3.3 Å$^3$. The discrepancy is due to the fact that in the method based on the Laguerre tessellation, there are no empty spaces within the protein, while it is possible that there are empty interstices amongst the atoms by using the geometrical model of WOLUMES.

It is not possible to compare the computational speed of these two methods, since VLDP is available as a web server. It is on the contrary possible to compare the computational speed of WOLUMES with that of another program, ATVOL, that computes atom volumes by separating neighbor atoms where their van der Waals spheres clashes and by performing Monte Carlo integrations to measure the volume of each atom (http://www.ctu.edu.vn/~dvxe/Bioinformatic/Software/kinemage/kinemage.biochem.duke.edu/software/software3.html#atvol). WOLUMES is clearly much faster, about 80 times more than ATVOL. The atom volumes computed with these two methods are quite similar (correlation coefficient = 0.803); the volumes computed with WOLUMES that tend to be slightly smaller than those computed with ATVOL. On average, the absolute value of the differences between the atom volumes computed with ATVOL and WOLUMES is equal to 3.13 (±0.02) Å$^3$. Such a discrepancy might be due to the different sets of van der Waals radii used by the two programs: while WOLUMES uses the set of radii determined by Li and Nussinov [21], ATVOL uses radii taken from older publications by Gavezzotti [24] and Bondi [25], which were not focused on protein molecules.

## 3 Atom and residue volumes

The same set of 501 protein structures described in the previous section was used to perform a basic statistical survey on atom and residue volumes in proteins.

The atom volumes are distributed around three frequency maxima (**Figure 3**). One, at bigger volumes, is due to the sulfur atoms of methionines and cysteines and to the carbon atoms of the methyl groups. Another maximum is due to the aliphatic side-chain carbon atoms bound to no more than two hydrogen atoms. And the third, at smaller volumes, is due to all the other carbon atoms and to all the nitrogen and oxygen atoms. The average values of the atom volumes are shown in **Table 4**. The atoms are grouped into the 24 types defined by Li & Nussinov [21]. As it is shown in **Figure 4**, the volumes of the isolated atoms, computed as the volume of a sphere of radius equal to the van der Waals radius of the atom, tend to be considerably larger that the volumes of the atoms located into the protein molecule. The relationship is V_WOLUMES = 18 + 0.5 V_ISOLATED (correlation coefficient = 0.763).

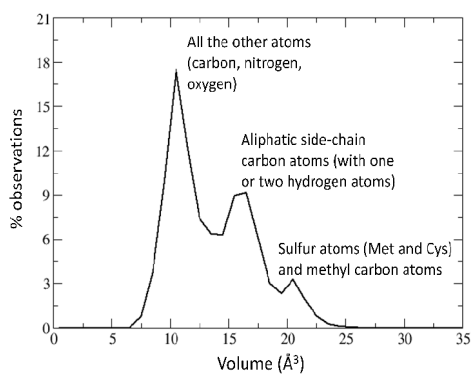

*Figure 3. Distribution of the atom volumes.*

It is not surprising that in proteins, all the types of atoms have a volume smaller that the volume that they would have if they were isolated, not connected covalently to other atoms, and not surrounded by other atoms non-bonded to them. In particular, the volumes of the Cα atoms and of the aromatic carbon atoms that are not bonded to hydrogen atoms experience a decrease of more than 60%, since they are equal to 11 and 9 Å$^3$ in the protein while they would be equal to 29 and 22 Å$^3$ if these atoms were isolated. The smallest contraction (30%) is exhibited by the methyl groups, by the side-chain sulfur atom of cysteine, and by the side-chain amino nitrogen atom of lysine, the volumes of which are equal to 21, 20, and 14 Å$^3$ in the proteins while they would be equal to 30, 28, and 20 Å$^3$ if these atoms were isolated from any other atom. On average, the isolated atoms would be spheres with a volume 43% larger than the volume they can occupy in proteins.

*Table 4. Average volumes of 24 atom types computed with the program WOLUMES on a set of 501 protein structures.*

| Symbol | Atom type | vdW radius (Å) | Volume (Å$^3$); standard error in parentheses |
|---|---|---|---|
| CA | Main chain alpha-carbon | 1.90 | 11.099(0.004) |
| C | Main chain carbonyl carbon | 1.75 | 10.336(0.002) |
| CH | Side-chain aliphatic carbon with one hydrogen | 2.01 | 15.496(0.005) |
| CH2 | Side-chain aliphatic carbon with two hydrogens, except those at beta-position and those next to a charged group | 1.92 | 16.533(0.004) |
| CH2b | Side-chain aliphatic carbon with two hydrogens at beta-position | 1.91 | 15.635(0.004) |
| CH2ch | Side-chain aliphatic carbon next to a charged group | 1.88 | 15.574(0.013) |
| CH3 | Side-chain aliphatic carbon with three hydrogens | 1.92 | 20.889(0.004) |
| CHar | Aromatic carbon with one hydrogen | 1.82 | 13.741(0.003) |
| Car | Aromatic carbon with no hydrogen | 1.74 | 8.579(0.004) |
| CHIm | Carbon on the imidazole side-chain of His | 1.74 | 12.705(0.010) |
| Cco | Side-chain carbonyl carbon | 1.81 | 11.587(0.007) |
| Ccoo | Side-chain carboxyl carbon | 1.76 | 10.872(0.005) |
| SH | sulfur on Cys | 1.88 | 20.284(0.045) |
| S | sulfur on Met | 1.94 | 20.202(0.018) |
| N | Main-chain amide nitrogen | 1.70 | 11.885(0.003) |
| NH | Side-chain nitrogen with one hydrogen | 1.66 | 10.053(0.013) |
| NH+ | Side-chain itrogen of His | 1.65 | 10.066(0.009) |
| NH2 | Side-chain neutral nitrogen with two hydrogen | 1.62 | 12.108(0.007) |
| NH2+ | Side-chain partially charged nitrogen on Arg | 1.67 | 11.776(0.014) |
| NH3+ | Side-chain nitrogen on Lys | 1.67 | 14.034(0.009) |
| O | Main-chain carbonyl oxygen | 1.49 | 8.261(0.003) |
| Oco | Side-chain carbonyl oxygen | 1.52 | 9.370(0.007) |
| Ocoo | Side-chain carboxyl oxygen | 1.49 | 9.092(0.004) |
| OH | Side-chain hydroxyl oxygen | 1.54 | 9.983(0.006) |

Some atoms are considerably larger than others. On average, they occupy 12.9 Å$^3$; however, the values are quite variable: some of them have an average volume larger than 20 Å$^3$ while the volumes of some other ones are smaller than 10 Å$^3$.

The average volumes of the 20 types of residues are given **Table 5**. They fit very well with the number of atoms contained in the residue (**Figure 5**), with a relationship that is Volume = 19.1 + 11.2 n_atoms (correlation coefficient = 0.952). This might seem to be surprising, since atom volumes are very variable. However, one must remember that several atom types (at least four) are present in all or in some of the amino acids, which then share many types of atoms.

The average volumes of the residues tend to be smaller than those computed by Tsai and colleagues with a Voronoi tessellation of the space [15]. However, this is not surprising since, as it was described above, space tessellation methods do not allow one to find many empty interstices inside the molecule, while the numerical procedure implemented in the program WOLUMES allow one to find empty regions in the protein core. The fact that this is the only difference between the volumes computed by Tsai and those computed with WOLUMES is demonstrated by their perfect correlation (V_WOLUMES = 13 + 0.67 V_Tsai; linear correlation coefficient = 0.996). Moreover, the presence of empty interstices into well packed materials is a well known chemical phenomenon, even in proteins [26], where non-functional small cavities can be filled, for example, by xenon atoms [27].

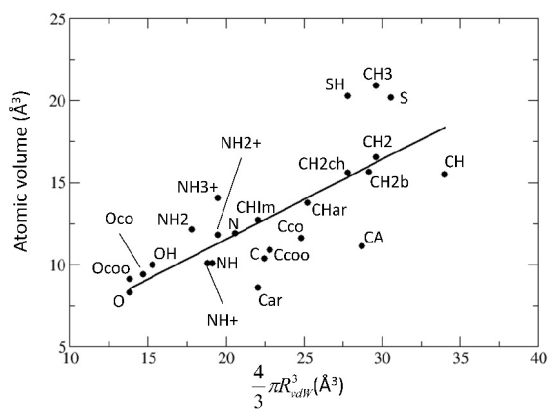

Figure 4. Relationship between the volumes of the isolated atoms and the volumes of the atoms in the protein.

Table 5. Average volumes of 20 residue types computed with the program WOLUMES on a set of 501 protein structures.

| Residue | Volume (Å$^3$); standard error in parentheses |
|---|---|
| Ala | 72.27(0.04) |
| Arg | 146.71(0.03) |
| Asn | 99.11(0.06) |
| Asp | 95.58(0.03) |
| Cys | 87.69(0.08) |
| Gln | 115.40(0.03) |
| Glu | 112.12(0.07) |
| Gly | 56.33(0.05) |
| His | 120.57(0.05) |
| Ile | 121.04(0.03) |
| Leu | 121.33(0.02) |
| Lys | 129.88(0.03) |
| Met | 126.38(0.07) |
| Phe | 142.42(0.03) |
| Pro | 96.14(0.03) |
| Ser | 78.64(0.04) |
| Thr | 96.01(0.11) |
| Trp | 171.16(0.06) |
| Tyr | 148.79(0.03) |
| Val | 104.88(0.03) |

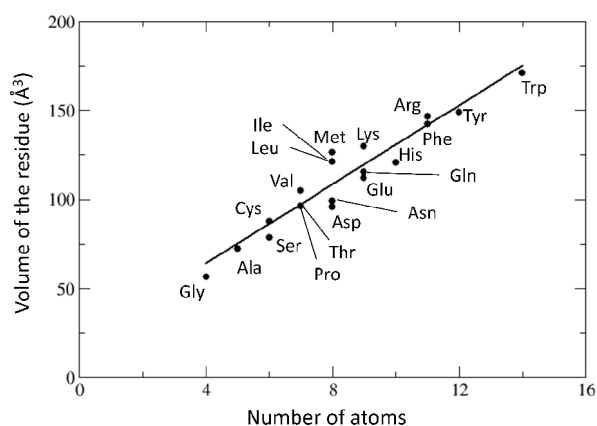

*Figure 5. Relationship between the volumes and the numbers of atoms of the twenty residue types.*

## 4 Conclusions

The program WOLUMES allows one to compute the volumes of atoms and residues in protein structures. It is a stand-alone program written in standard C and it can be used, as a consequence, on any operating system. Its results are comparable to those of other programs, though it is considerably faster. It does not depend on the licences of other programs and is thus also of easy installation on any computer. The fact that it is stand-alone and fast is important for both large scale proteomic applications, where tens of thousands of protein structures are processed, and applications to single proteins, where the structural biologists may need to examine manually the structures, avoiding unnecessary complications of more complicated software packages.

The program WOLUMES uses the set of van der Waals radii determined by Li and Nussinov [21] on the basis of the analysis of contact distance distributions. Other values can easily be introduced by the user in the file 'WOLUMES.vdw', like for example the radii determined by Tsai and colleagues on the basis of contact distances in small molecule structures [15] or other, older radii sets [28, 29].

The programs requires also a database of atom and residues volumes extracted from a set of monomeric protein structures, which is used to identify the atoms and the residues that are unusually very small or large (the smallest and largest 1% of the observations). It provides two output files: WOLUMES.res, with the volumes of the residues and WOLUMES.ato. with the volumes of the atoms, written in the PDB format with the atom volume in the field of the B-factor. Detailed instructions on how to run the software are provided with the source code and some examples.

### Acknowledgements

I thank Kristina Djinovic for helpful discussions. This work was partially made possible by the BIN-III programme within the Austrian GEN-AU initiative.